\documentclass[aps,prc,superscriptaddress,reprint,nofootinbib]{revtex4-1}

\usepackage{graphicx}
\usepackage{multirow}
\usepackage{color}

\usepackage{amssymb}
\usepackage{bbm}
\usepackage{amsmath}
\usepackage{color}
\usepackage{lineno}
\usepackage{enumitem}
\usepackage{scalerel,stackengine}
\usepackage{comment}
\usepackage[normalem]{ulem}
\usepackage{diagbox}

\stackMath
\newcommand\reallywidehat[1]{%
\savestack{\tmpbox}{\stretchto{%
  \scaleto{%
    \scalerel*[\widthof{\ensuremath{#1}}]{\kern-.6pt\bigwedge\kern-.6pt}%
    {\rule[-\textheight/2]{1ex}{\textheight}}
  }{\textheight}%
}{0.5ex}}%
\stackon[1pt]{#1}{\tmpbox}%
}

\begin{document}

\title{Testing the validity of the surface approximation for reactions induced by weakly bound nuclei with a fully quantum-mechanical model}


\author{Junzhe Liu}

\affiliation{School of Physics Science and Engineering, Tongji University, Shanghai 200092, China.}

\author{Jin Lei}
\email[]{jinl@tongji.edu.cn}

\affiliation{School of Physics Science and Engineering, Tongji University, Shanghai 200092, China.}

\author{Zhongzhou Ren}

\affiliation{School of Physics Science and Engineering, Tongji University, Shanghai 200092, China.}


\begin{abstract}
We examine the validity of surface approximation for breakup reactions using a fully quantum-mechanical model proposed by Ichimura, Austern, and Vincent (IAV). Analogous to the semi-classical picture, we introduce radial cut-offs to scattering waves in the IAV framework, which we refer to as IAV-cut.  Systematic calculations are conducted for nonelastic breakup reactions induced by $^6$Li and deuterons at various incident energies. A comparison between the results obtained from IAV and IAV-cut is performed. The excellent agreement observed between IAV and IAV-cut in $^{6}$Li induced reactions, regardless of incident energy and target nuclei, signifies their insensitivity to the inner part of the scattering wave function, thus providing validation for the semi-classical picture. However, for deuteron induced breakup reactions, the IAV-cut results exhibit a suppression in the cross section, suggesting a strong dependence on the interior wave functions. This suppression is further enhanced as the incident energy increases.
\end{abstract}

\date{\today}%
\maketitle

\section{Introduction \label{sec:intro}}
The breakup of a nucleus into two or more fragments
is an important mechanism among various channels of nuclear reactions. With the recent advancements in radioactive beam facilities, measuring the breakup reactions of rare atomic nuclei is now feasible~\cite{abel2019isotope}. This development has greatly enhanced our understanding of nuclear properties such as binding energy, spectroscopic factors, and angular momentum~\cite{PhysRevLett.103.262501}.
Presently, coupled-channel methods which can deal with the excitation of the internal freedom has been widely used to calculate the cross section of rare nuclei induced reactions~\cite{HAGINO2022103951}.

In some experiments, weakly bound  nuclei are produced to bombard with a target, ultimately fragmenting into two separate components. 
From an experimental perspective, determining all particles simultaneously and specifying the  final states of each fragment are challenging. 
Alternatively, if the experiment is designed to detect only one of the fragments inclusively, the process can be simplified to $a + A \rightarrow b + B^*$, where the projectile $a$ is assumed to have a two-body structure $(b+x)$ and $B^*$ represents any possible state of the $x + A$ system. This process of inclusive breakup has been extensively studied in experimental research~\cite{Duan22,Wang21,yang21,DiPietro19,Duan20}.
If the three particles, $b$, $x$, and $A$, remain in their ground state after the breakup, the corresponding process is referred to as elastic breakup (EBU). Breakup accompanied by target excitation, fusion between $x$ and $A$, and any possible mass rearrangement between $x$ and $A$ is referred to as nonelastic breakup (NEB). Precise calculations for NEB are necessary, for example, to exam the semi-classical approach which has been widely applied to knockout reaction~\cite{AUMANN2021103847}, and in the surrogate method applied to study nuclear synthesis and the chemical evolution of stars~\cite{RevModPhys.84.353}. Therefore, the evaluation of NEB cross sections is of great value both theoretically and experimentally.

In 1985, Hussein and McVoy (HM) derived one of the earliest closed-form formulae for the inclusive breakup cross section~\cite{HUSSEIN1985124}. HM's derivation provided deep insight through the summation over all $x$-$A$ states. By utilizing the Glauber approximation to analyze scattering waves, they obtained an appealing and intuitive form with a clear probability interpretation of the breakup reaction. This evaluation of the NEB cross section is exclusively dependent on the asymptotic properties ($S$-matrix) between the fragments ($b$ or $x$) and the target. This is a consequence of employing the semi-classical Glauber approximation. The HM model, along with structure calculations, finds extensive application in spectroscopic studies of one-nucleon removal reactions~\cite{TOSTEVIN2001320,PhysRevC.90.057602,AUMANN2021103847}.
Furthermore, D. Baye and colleagues developed the dynamical eikonal model to address dissociation cross sections~\cite{PhysRevLett.95.082502}. Rather than employing the adiabatic approximation used in the standard eikonal model for phase shift evaluation, they numerically solve a semi-classical time-dependent Schrödinger equation using straight-line trajectories. This model has found application in investigating reactions involving halo nuclei~\cite{PhysRevC.81.024606,PhysRevC.70.064605}.

The transfer to continuum (TC) model is another successful semi-classical approach used to evaluate the NEB cross section~\cite{PhysRevC.38.1776}. In the TC model, the transfer amplitude between the initial and final states is calculated using a time-dependent approach~\cite{PhysRevC.63.044604}. This transfer amplitude is calculated by using the asymptotic part of the initial bound state and the final continuum state. The main principle of this semi-classical approximation is to utilize the classical trajectory for approximating the relative motion between the projectile and the target. This semi-classical TC method has been widely applied to large numbers of break up reactions, from stable to exotic projectiles~\cite{PhysRevC.44.1559,BONACCORSO20181}.

In spite of the tremendous success attained by the previously mentioned semi-classical models, research has already been conducted to establish a quantum-mechanical model. In the early 1980s, Udagawa and Tamura (UT)~\cite{PhysRevC.24.1348}  developed their NEB formalism using DWBA, while Austern and Vincent (AV)~\cite{PhysRevC.23.1847} carried out a similar derivation. After a long-standing dispute between these two groups, the equivalence of these two derivations has finally been proved in Ichimura, Austern, and Vincent (IAV)'s work~\cite{PhysRevC.32.431}. Due to the computational limitations, this model is not implemented numerically until recently~\cite{PhysRevC.92.044616,Potel15,Carlson2016}, and its validity has finally been tested through numbers of applications~\cite{PhysRevLett.122.042503,PhysRevLett.123.232501}.
This fully quantum-mechanical model starts from the effective three-body Hamiltonian, making no assumptions on the trajectory, and maintain the conservation laws naturally.
The EBU is a process that all fragments and targets are properly separated, which allows us to assume that the cross section of this process depends solely on the asymptotic part of wave functions. Therefore, the EBU cross sections are unaffected by the interior part of wave functions~\cite{baur1986breakup}. 
Nevertheless, this conclusion may not apply to NEB, because NEB contains the fusion channel between the fragments and the target, and the calculation requires a short-range imaginary part of the optical potential to describe this fusion process.
However, HM model with Glauber approximation ignores the inner part of the scattering wave function due to the semi-classical approximation, which lacks a direct comparison to fully quantum-mechanical approaches. Here we present a study on this surface approximation\footnote{In this study, we refer to surface approximation as a procedure that relies on the asymptotic properties of the scattering wave function.}, by introducing a radial cut-off to the scattering functions, where no other semi-classical assumptions need to be taken. We refer to this cut-off method in the IAV framework as IAV-cut. By varying the cut-off radius, we  study the sensitivity to the inner wave functions, and thus test the validity of these semi-classical interpretations of reaction processes. 

The paper is organized as follows. In Sec.~\ref{sec:theory} we review the formalism of the IAV model and raise our surface approximation through the radial cut-off. In Sec.~\ref{sec:app} we apply this cut-off to several inclusive reactions induced by $^{6}$Li and deuterons. Finally, in Sec.~\ref{con} we summarize the main results of this work and outline some future developments. 

\section{\label{sec:theory}Theoretical framework}
In this section, we briefly review the IAV model~\cite{AUSTERN1987125,PhysRevC.32.431} and define our corresponding surface approximation in the IAV model.

The inclusive breakup reaction under study takes the form
\begin{equation}
    a(=b+x)+A \rightarrow b+ B^{*},
    \label{eq1}
\end{equation}
where the projectile $a$ has a two body structure $(b+x)$, $b$ is the detected particle, and $B^{*}$ denotes any possible final state of the $x+A$ system.
In  IAV model, fragment $b$ is called the spectator, and fragment $x$ is called the participant.

The IAV model gives the NEB cross section
\begin{equation}
    \left.\frac{\mathrm{d}^2\sigma}{\mathrm{d}\Omega_b \mathrm{d}E_b}\right|_{\textbf{post}}^{\textbf{NEB}}
    =
    -\frac{2}{\hbar v_a}\rho_b(E_b) 
    \langle 
    \psi_x (\boldsymbol{k_b})|
    W_x 
    |\psi_x (\boldsymbol{k_b})
    \rangle,
    \label{eq:IAV}
\end{equation}
where 
$v_a$ is the projectile-target relative velocity, 
$\rho_b(E_b)=\mu_b k_b/[(2\pi)^3\hbar^2]$ is the density of states for particle $b$, $\mu_b$ and $k_b$ are the reduced mass and wave number, respectively, 
$W_x$ is the imaginary part of $U_{x}$ which describes $x$+$A$ elastic scattering, 
$\psi_x$ is the so-called $x$-channel wave function which is obtained by solving the inhomogeneous differential equation
\begin{equation}
    (E_x-K_x-U_{x})\psi_x(\boldsymbol{k_b},\boldsymbol{r_x})
    =
    \langle \boldsymbol{r_x}
    \chi_b^{(-)}(\boldsymbol{k_b})
    |V_{\mathrm{post}}|
    \chi_a^{(+)}\phi_a\rangle,
    \label{eq:inhomo}
\end{equation}
where $E_x=E-E_b$, 
$K_x$ is the kinetic energy operator for relative motion between fragment $x$ and target $A$,
$\chi_b^{(-)}$ is the scattering wave function with incoming boundary condition describing the scattering of $b$ in the final channel with respect to the $x$+$A$ subsystem, $V_{\mathrm{post}}=V_{bx}+U_{bA}-U_{bB}$ is the post form transition operator, where $V_{bx}$ is the potential binding two clusters $b$ and $x$ in the initial composite nucleus $a$, $U_{bA}$ is the fragment-target optical potential, $U_{bB}$ is the
optical potential in the final channel,
$\chi_a^{(+)}$ is the distorted-wave describing the $a$+$A$ elastic scattering with an outgoing boundary condition, and $\phi_a$ is the initial ground-state of the projectile $a$.
To simplify the calculations, we
ignore intrinsic spins. 
As for the angle integrated NEB cross section, we have the partial wave expansion form of Eq.~(\ref{eq:IAV}),
\begin{equation}
\begin{aligned}
    \left. \frac{\mathrm{d}\sigma}{\mathrm{d}E_b}\right|_{\textbf{post}}^{\textbf{NEB}}
    =
    &-\frac{1}{2\pi\hbar v_a}\rho_b(E_b) 
    \frac{1}{2l_{bx}+1}\\
    &\times
   \sum_{l_al_bl_x} \int \mathrm{d}r_x r_x^2|\mathcal{R}_{l_al_bl_x}(r_x)|^2 W_x(r_x), 
\end{aligned}
\label{eq:expansion}
\end{equation}
where $\mathcal{R}_{l_a l_b l_x}$ represents the radial part in the partial wave expansion of $\psi_x$. The variables $l_a$, $l_b$, $l_{bx}$ and $l_x$ represent the relative angular momenta between $a$ and $A$, $b$ and $B^*$, $b$ and $x$, and $x$ and $A$, respectively. Specifically, $l_{bx}$ is determined by the initial bound state of the projectile, while the maximum values of $l_a$, $l_b$, and $l_x$ are selected to ensure the convergence of the cross section.
More details of the IAV model can be found in Ref.~\cite{PhysRevC.92.044616} and its Appendix.

When the Coulomb interaction is taken into consideration, the incoming and outgoing distorted wave have the partial wave expansions,
\begin{equation}
    \langle r_a l_a m_a |\chi_a^{(+)} (\boldsymbol{k_a})\rangle
    =
    \frac{4\pi}{k_a r_a} i^{l_a}e^{i\sigma_{l_a}}
    u_{l_a}(r_a)
    \left [Y_{l_a}^{m_a}(\hat{k_a})\right]^{*},
\end{equation}
\begin{equation}
    \langle \chi_b^{(-)}(\boldsymbol{k_b}) |r_b l_b m_b \rangle
    =
    \frac{4\pi}{k_b r_b} i^{-l_b}e^{i\sigma_{l_b}}
    u_{l_b}(r_b)
    Y_{l_b}^{m_b}(\hat{k_b}),
\end{equation}
where $k_a$ and $k_b$ are the relative wave numbers of the incident and outgoing channel, $\sigma_{l_a}$ and $\sigma_{l_b}$ are the Coulomb phase shift. One important numerical task is to determine the radial wave function $u_{l_a}$($u_{l_b}$), and our surface approximation method focus on the cut-off of these wave functions. In particular, we set the radial wave function $u_{l_a}$($u_{l_b}$) to zero below a specific cut-off radius. 

The surface approximation in nuclear reactions often means using the asymptotic behavior of the wave function to calculate the cross section~\cite{PhysRevC.8.1084,kasano1982new}. 
Instead of solving the radial Schrödinger equations, others use eikonal method to calculate the phase shift ~\cite{HUSSEIN1985124}.  According to the unitarity of $S$-matrix, the cross section of NEB, which represents the absorption of participant $x$ by target $A$, can be expressed using $S$-matrices~\cite{pampus1978inclusive}.  The key point of this kind of surface approximation is to extract the reaction information from the asymptotic behavior ($S$-matrix). In another word, only the exterior part of the scattering wave function influences the cross sections. 
However, in the IAV model, the key step is to solve the differential inhomogeneous equation to obtain the $x$-channel wave function. This wave function is subsequently used for computing the NEB cross section.
As suggested by Baur~\cite{baur1986breakup}, one suitable surface approximation for the IAV model may be 
replacing the wave function with some suitable form. The simplest one is the asymptotic form
\begin{equation}
    u_l(r)
    \approx
    \frac{i}{2}\left [ H^{(-)}_l(r)-S_lH^{(+)}_l(r) \right ].
\end{equation}
However, due to the irregularity of this asymptotic form at the origin, it can not be implemented numerically.
In order to prevent the divergence at the origin and maintain the boundary condition of wave function, we introduce a radial cut-off to the scattering wave functions both in the entrance and the exit channel consistently, which is mentioned by Baur~\cite{baur1986breakup} as well,

\begin{equation}
    u_{l_a}(r)=u_{l_b}(r)=0   \qquad r<R_{\mathrm{cut}},
\end{equation}
where $R_{\mathrm{cut}}$ is the cut-off radius which is chosen according to the interaction radius of optical potential.
It is important to note that implementing this cut-off will result in a discontinuity in the wave functions $u_{l_a}(r)$ and $u_{l_b}(r)$ at the cut-off radius.
In the subsequent discussion, we will refer to the calculation using this cut-off method as IAV-cut.
 
The angular integrated NEB cross section can be directly obtained by using the radial component of the $x-$channel wave function, $\mathcal{R}_{l_a l_b l_x}$. Therefore, it is crucial to investigate the impact of the cut-off, especially when summing over $l_b$ and $l_x$ and only retaining the dependence on $l_a$, which represents the angular momentum between the projectile and target. We denote this new radial part wave function as $R_{l_a}(r_x)$, and its modulus square has the relation to $\mathcal{R}_{l_al_bl_x}(r_x)$
\begin{equation}
\label{eq:rla}
    |R_{l_a}(r_x)|^2 = \sum_{l_b l_x}|\mathcal{R}_{l_al_bl_x}(r_x)|^2.
\end{equation}
Then the angular integrated NEB cross section can be obtained by 
\begin{equation}
\begin{aligned}
    \left. \frac{\mathrm{d}\sigma}{\mathrm{d}E_b}\right|_{\textbf{post}}^{\textbf{NEB}}
    =
    &-\frac{1}{2\pi\hbar v_a}\rho_b(E_b) 
    \frac{1}{2l_{bx}+1}\\
    &\times
   \sum_{l_a} \int \mathrm{d}r_x r_x^2|R_{l_a}(r_x)|^2 W_x(r_x), 
\end{aligned}
\label{eq:expansion1}
\end{equation}

\section{\label{sec:app}Application}

In this section, we present systematic calculations for the inclusive breakup induced by $^6\mathrm{Li}$ and deuteron projectiles and compare the results of IAV with IAV-cut.
The choice of cut-off radii is critical for our implementation. A cut-off radius that is too small will have little impact due to the removal of only a small part of the wave function. However, a cut-off radius that is too large will obscure the interacting details between the two nuclei, leading to a significant decrease in the cross section.
In the global optical potential model~\cite{COOK1982153,PhysRevC.9.2010,PhysRevC.74.044615,koning2003local}, the radius parameter that we used in the current study takes the form
\begin{equation}
    R_0=r_{0}\times A_T^{1/3},
\end{equation}
where $r_0$ is the geometric parameter of optical potential, $A_T$ is the mass number of the target. 
The parameter $R_0$ represents the effective range of the nuclear force. The mass number of the projectile is often omitted in the fitting of the global optical potential. Consequently, we select the cut-off radius to be of a similar order of magnitude as $R_0$. This selection of the cut-off radius accounts for the variability in the effective range of interaction across different target nuclei.

\subsection{Convergence of the numerical method}
As previously mentioned, introducing a radial cut-off for the wave function leads to a discontinuity at the cut-off radius $R_{\mathrm{cut}}$. In numerical calculations, the Gaussian quadrature method is widely used to save computing time by integrating the wave function. However, accurately capturing this discontinuity at the cut-off radius $R_{\mathrm{cut}}$ often requires additional integration quadrature points. Consequently, using an excessive number of grid points in the integration significantly increases computer memory usage and computation time. Moreover, the Gaussian quadrature method is characterized by having more grid points at the upper and lower limits of the integration interval compared to the equally spaced grid points of Simpson's rule and the trapezoidal rule. 
As a result of implementing a radial cut-off for the wave function, somes quadrature points near the origin in the Gaussian method do not contribute to the overall integration result since the integrand becomes zero due to the cut-off.
This leads to the waste of our computational resources. To improve the numerical efficiency, when evaluating the source term, which is the inhomogeneous term on the right-hand side of Eq.~(\ref{eq:inhomo}), we choose the integral\footnote{More details of the evaluation of this source term can be seen in the Eq.~(13) of Ref.~\cite{PhysRevC.97.034628} and its appendix.
The variable $r_{bx}$ in the Eq.~(13) of Ref.~\cite{PhysRevC.97.034628} is expressed by the coordinate set $(r_x, r_b)$ as presented in Eq. (A5) in the appendix. The choice of $r_b$ starting from $R_{\mathrm{cut}}$ will break the continuity of wave function in $r_{bx}$ as well, so a test of convergence is necessary. } region to start from $R_{\mathrm{cut}}$, rather than setting the wave functions to zero and carrying out the integral from the origin. By reselecting the integration interval in this way, we no longer calculate the parts that do not contribute to the cross section, thus achieving rapid convergence of the integration result with fewer integration points.

\begin{figure}[tb]
    \begin{center}
  {\centering \resizebox*{1.0\columnwidth}{!}{\includegraphics{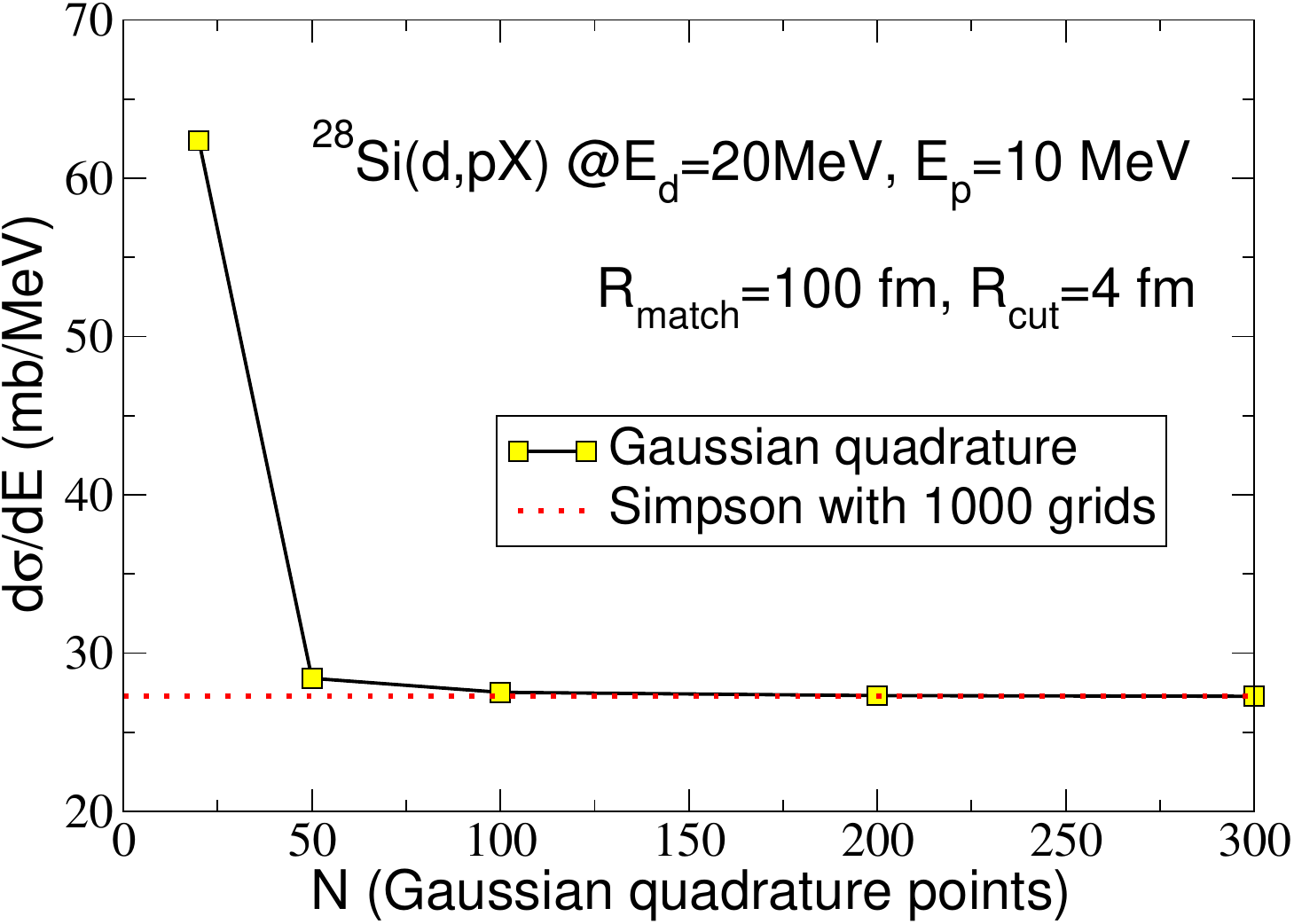}} \par}
\caption
{NEB cross section of $^{28}\mathrm{Si}(d,p X)$ at $E_d=30$ MeV with relative outgoing energy $E_p=10$ MeV, as a function of numbers of Gaussian quadrature points.
The dotted horizontal line represent the cross section with 1000 Simpson's grids.}
 \label{converge}
 \end{center}
 \end{figure}

To test the validity of this method, we consider the $^{28}\mathrm{Si}(d,p X)$ reaction at the incident kinetic energy of 30 MeV in the lab frame and relative kinetic energy between $p$ and $^{29}\mathrm{Si}^*$ of 10 MeV in the CM frame. We plotted the differential cross section $\mathrm{d}\sigma/\mathrm{d}E$ in Fig.~\ref{converge} as a function of the number of Gaussian quadrature points. The figure shows that increasing the number of quadrature points from 25 to 50 results in a sharp drop in the cross section, but further increasing the number of points leads to little change in the cross section and good convergence. Our test also demonstrated that to achieve the same convergence using Simpson's rule (shown as the dotted horizontal line), we needed to employ a minimum of 1000 grids, which is significantly greater than the number of quadrature points ($>$100) required for the Gaussian method. The good convergence shown in the figure supports our choice of the integral region. Similar rapid convergence can be achieved in other reaction systems as well.
\subsection{Application to $(^6\mathrm{Li},\alpha X)$}
Since the Glauber approximation is widely used in the heavy ion induced knockout reactions~\cite{doi:10.1146/annurev.nucl.53.041002.110406,TOSTEVIN2001320}, we present studies on these reactions with IAV-cut to test the validity of the surface approximation.
Here we consider the calculations for the $^{208}\mathrm{Pb}(^6\mathrm{Li},\alpha X)$ reaction.
We treat $^6$Li as $\alpha$+$d$ cluster in the following discussion.
The incoming channel optical potential, which describes the $^6\mathrm{Li}$+$^{208}$Pb elastic scattering, is taken from Ref.~\cite{COOK1982153}. Besides, the $\alpha$+$
^{210}$Bi$^{*}$ and $\alpha$+$^{208}\mathrm{Pb}$ interaction are adopted from Ref.~\cite{PhysRevC.9.2010}, and the $d$+$^{208}\mathrm{Pb}$ interaction is taken from Ref.~\cite{PhysRevC.74.044615}.
The potential binding fragments $\alpha$ and $d$ in the initial composite projectile is assumed to take the Woods-Saxon (WS) form with the following parameter set: $a_v=0.7$ fm and $r_v=1.15$ fm. The depth of this WS potential is fitted to reproduce the experimental binding energy of $^6\mathrm{Li}$.

\begin{figure}[tb]
    \begin{center}
  {\centering \resizebox*{1.0\columnwidth}{!}{\includegraphics{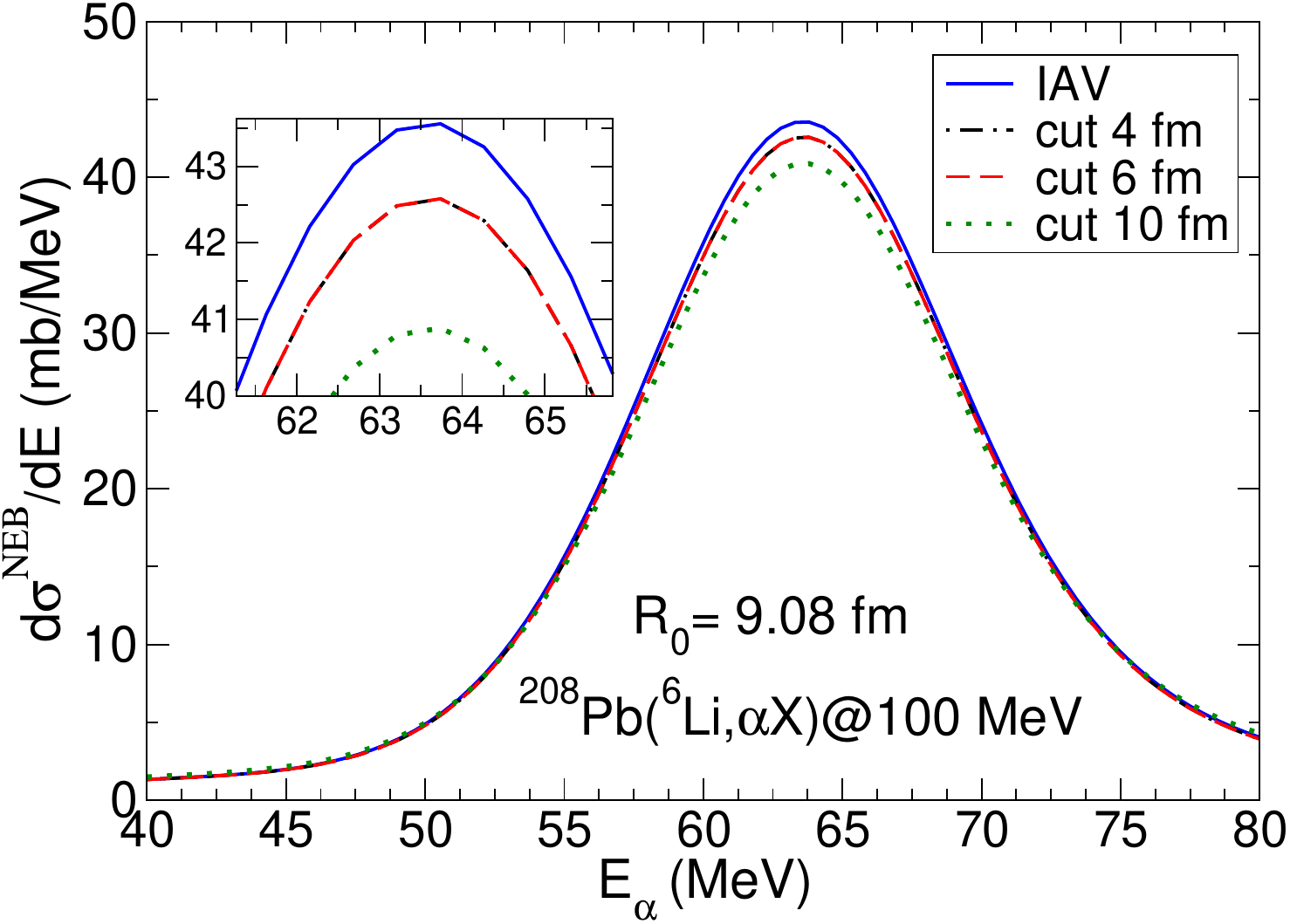}} \par}
\caption
{The angular integrated differential NEB cross section of $^{208}$Pb($^6$Li, $\alpha X$) as a function of the outgoing energy $E_{\alpha}$ in the CM frame, at a laboratory energy of 100 MeV. The solid line represents the result from the IAV model, whereas the dot-dashed, dashed, and dotted lines correspond to the cases with cut-off radii of 6 fm and 10 fm for IAV-cut, respectively.}
 \label{dsde}
 \end{center}
 \end{figure}

The nominal
Coulomb barrier for this system is around $30.1$ MeV~\cite{PhysRevC.66.041602}.
The model space needed for converged solutions of the
IAV model contains partial waves $l\le90$ in the $^6\mathrm{Li}$+$^{208}\mathrm{Pb}$ and $\alpha$+$^{210}\mathrm{Bi}^{*}$ relative
motion, and $l\le40$ in the $d$+$^{208}\mathrm{Pb}$ channel at $E_{\mathrm{lab}}=100$ MeV.
The model space we chose was large enough to ensure the convergence of NEB cross section. 
For the $^{208}\mathrm{Pb}(^6\mathrm{Li},\alpha X)$ reaction, the radius parameter of the imaginary part of the optical potential between $^6$Li and $^{208}$Pb is $R_0=9.08$ fm~\cite{COOK1982153}, so we choose the cut-off radius to be 4 fm, 6 fm, and 10 fm according to the previous discussion on the selection of cut-off parameter. It is important to note that a cut-off is applied consistently to both the incoming channel scattering wave function of $^6$Li+$^{208}$Pb and the outgoing channel scattering wave function of $\alpha$+$^{210}$Bi$^{*}$.
The differential cross section of this system at $E_{\mathrm{lab}}=100$ MeV as a function of the  outgoing  energy of $\alpha$ particles in CM frame is presented in Fig.~\ref{dsde}.
The solid line corresponds to results from the IAV model, while the dot-dashed, dashed, and dotted lines represent the cases for IAV-cut with cut-off radii of 4 fm, 6 fm, and 10 fm, respectively. First we notice that the four curves share the same shape, and the peaks are located around the same outgoing energy. 
We observe that for outgoing kinetic energies lower than 50 MeV or higher than 75 MeV, the four curves almost overlap, suggesting that the effect of the cut-off radius is minimal.
However, between 50 MeV and 75 MeV, an increasing difference can be observed as the cut-off radius increases. Nevertheless, even in the worst case (i.e., with a 10 fm cut-off), the difference is still smaller than the typical experimental uncertainty.
Interestingly, we can also see that the difference between the 4 fm cut and 6 fm cut cases is very small and almost invisible in the figure. This indicates that the corresponding part from 4 to 6 fm of the wave function does not affect the cross section significantly.
These results suggest that IAV-cut produces satisfactory outcomes compared to the original IAV calculation. 

\begin{figure}[h]
    \begin{center}
  {\centering \resizebox*{1.0\columnwidth}{!}{\includegraphics{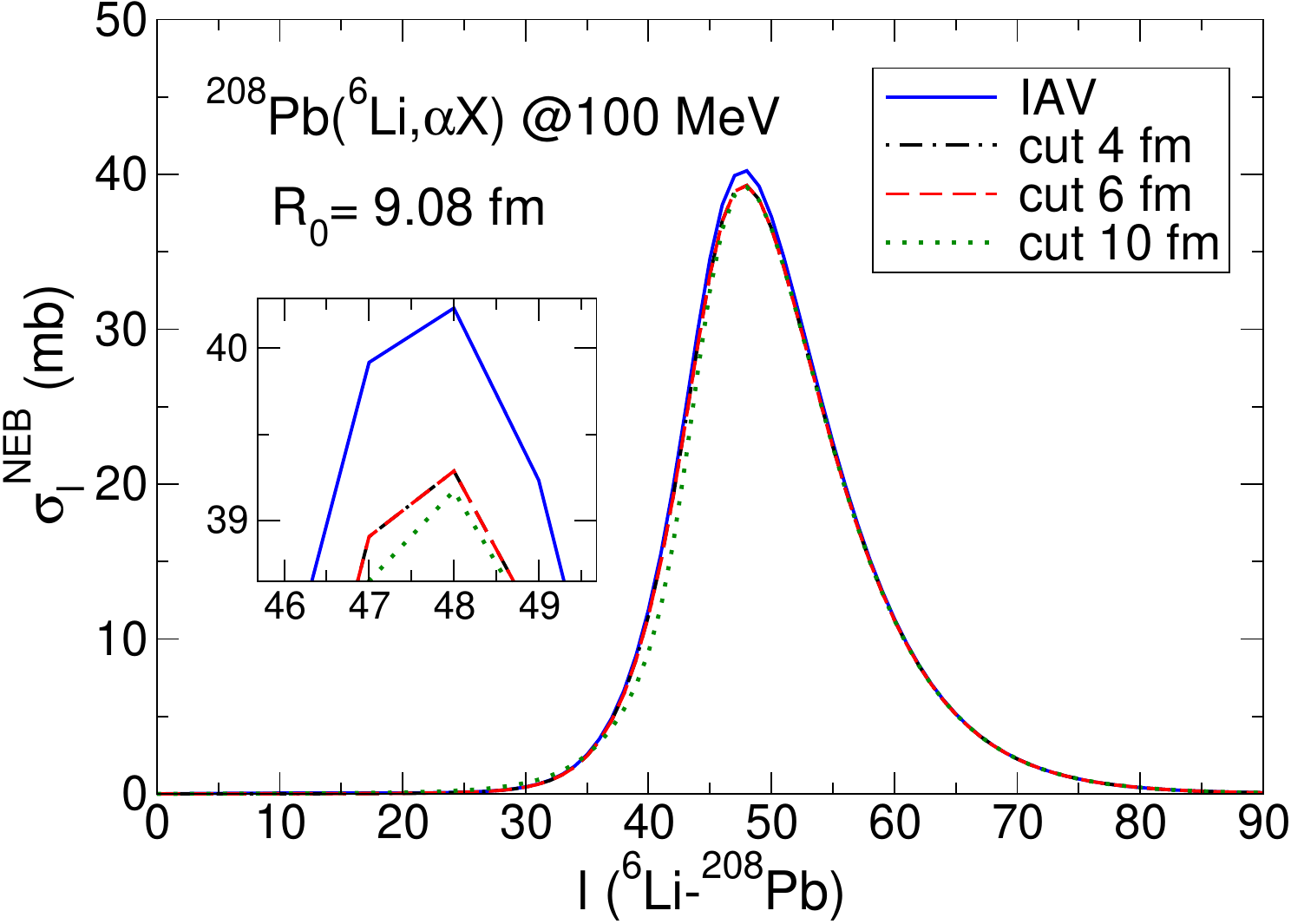}} \par}
 \caption{Integrated \sout{$\alpha$}  NEB cross section, as a function of the
relative angular momentum between $^6\mathrm{Li}$ and $^{208}\mathrm{Pb}$, for the $^{208}\mathrm{Pb}(^6\mathrm{Li},\alpha X)$ reaction at $E_{\mathrm{lab}}=100$ MeV. The solid line corresponds to results from the IAV model, while the dot-dashed, dashed, and dotted lines represent the cases for IAV-cut with cut-off radii of 4 fm, 6 fm, and 10 fm, respectively.}
 \label{sla100}
 \end{center}
 \end{figure}

\begin{figure}[h]
  \includegraphics[width=0.48\textwidth]{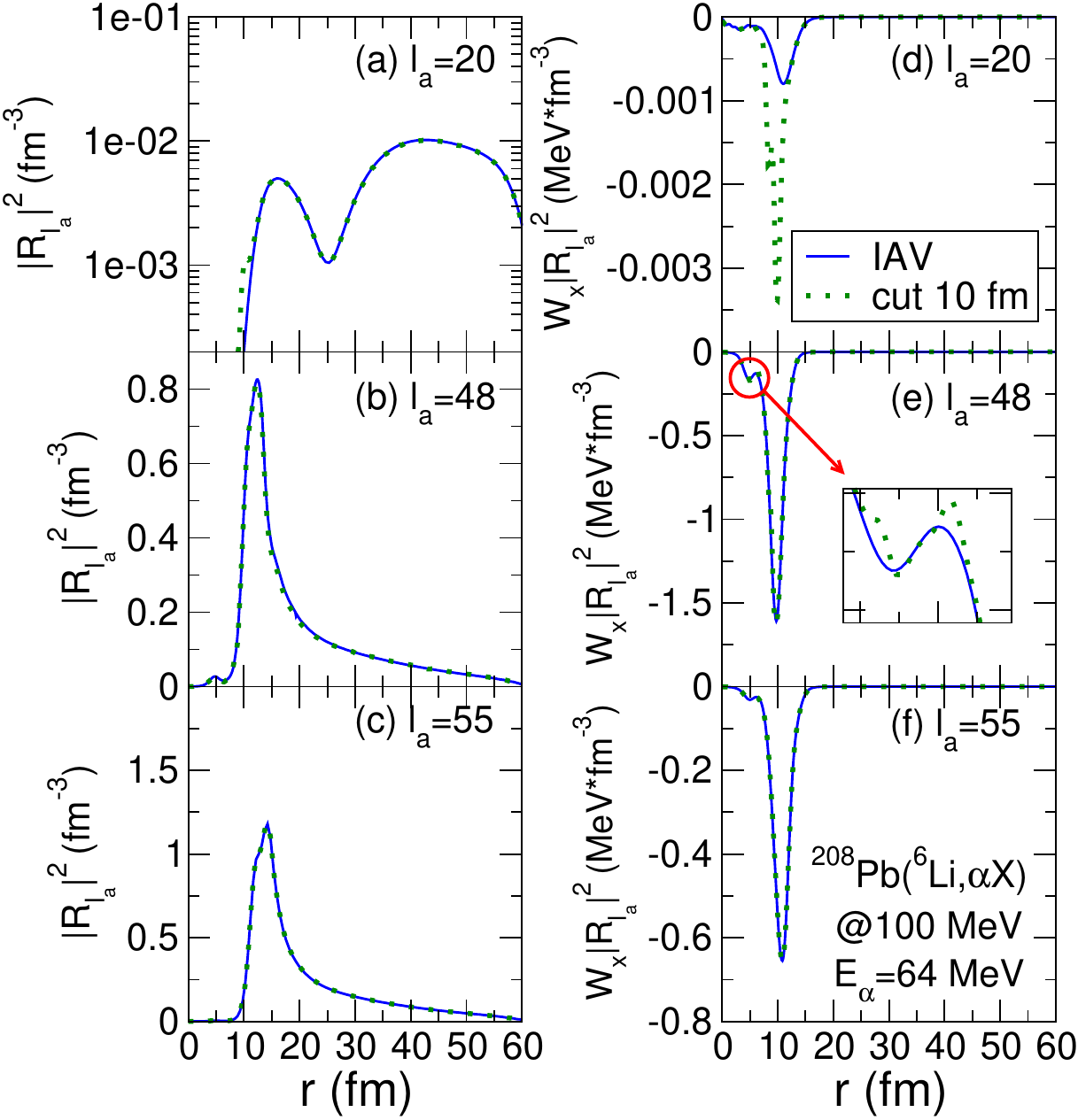}
\caption{Modulus square of the radial part of $d$+$^{208}\mathrm{Pb}$ wave function defined in Eq.~(\ref{eq:rla}) for (a) $l_a=20$, (b) $l_a=48$, and (c) $l_a=55$, for the $^{208}\mathrm{Pb}(^6\mathrm{Li},\alpha X)$ reaction at $E_{\mathrm{lab}}=100$ MeV and relative outgoing kinetic energy $E_{\alpha}=64$ MeV. The solid line represents the IAV result, while the dotted line corresponds to the IAV-cut case with 10 fm cut-off. 
The product of the imaginary part of the $d$+$^{208}\mathrm{Pb}$ potential and the wave function for (d) $l_a=20$, (e) $l_a=48$, and (f) $l_a=55$.}
\label{LiR2}
\end{figure}

Given the substantial importance of the angular momentum dependency of cross sections, we proceeded to examine the partial wave distribution of the cross sections and the effects of this cut-off method on the cross sections.
The projectile-target angular momentum distribution of integrated NEB cross section for the same reaction is shown in Fig.~\ref{sla100}.
It is observed that the peaks are determined at approximately the same value of $l_a$ which stands for the relative angular momentum between $^{6}\mathrm{Li}$ and ${^{208}\mathrm{Pb}}$, and all curves exhibit the same bell-shaped distribution as described in~\cite{lei2021comparison}.
The cross section shows a strong absorption effect of the $^6$Li+$^{208}$Pb interaction for low partial waves ($l_a\leq20$), thus leading to zero NEB cross section.
Furthermore, the impact of the cut-off is only evident for partial waves within the range $20\leq l_a\leq55$, with higher partial waves remaining unaffected.

Mathematically, wave functions corresponding to large angular momentum states are equal to zero at sufficiently small radii because it is hard to penetrate into the high centrifugal barrier. As a result, setting the inner part of these wave functions to zero will not change the calculation of the cross section.
These results illustrate a consistent agreement between cases using various radial cut-offs and the direct calculation based on the IAV model. This confirmation validates the utilization of the surface approximation in this reaction. In this context, the term surface approximation implies that the NEB cross section remains unaffected by the inner part of the incoming scattering wave of $^6$Li+$^{208}$Pb and the outgoing scattering wave of $\alpha$+$^{210}$Bi$^{*}$.

We pick out the $l_a=$ 20, 48, and 55 cases and draw $|R_{l_a}|^2$ at relative outgoing kinetic energy $E_{\mathrm{\alpha}}=64$ MeV in CM frame in Figs.~\ref{LiR2} (a), (b) and (c), respectively. The partial waves with $l_a=20$, 48, and 55 belong to distinct regions as discussed in Fig. 3: strong absorption where the NEB cross section is close to zero, the maximum value of $\sigma_l$ where the difference between the results obtained by the IAV model and IAV-cut is significant, and the region where the centrifugal barrier plays an influential role. The solid lines represent the IAV results, while the dotted lines denote the IAV-cut results with 10 fm cut-offs. 
First, the difference between wave functions shown in Fig.~\ref{LiR2}(a) does not exhibit a significant difference.
Due to the strong absorption effect of the $^6$Li+$^{208}$Pb interaction, the probability flux for the low angular momentum component is removed to the fusion channel, resulting in a relatively small wave function for the breakup process. Thus, this low partial wave makes a minimal contribution to the NEB cross section. As for the $l_a=48$ case shown in Fig.~\ref{LiR2}(b), 
a small difference occurs in the range of 15 $\sim$ 25 fm and near 5 fm,  while leaving its asymptotic parts unaltered. Since the EBU cross section is evaluated via the boundary conditions ($S$-matrices)~\cite{pampus1978inclusive}, this surface approximation is valid for EBU calculations as well.
Additionally, as shown in Fig.~\ref{LiR2}(c), no apparent difference appears in the wave functions at $l_a=55$. 
This finding is consistent with the results in Fig.~\ref{sla100}, which also does not exhibit clear changes in the high partial wave component.

As discussed in Eq.~(\ref{eq:expansion1}), NEB cross sections can be evaluated by  $|R_{l_a}|^2W_x$, which is the product of the modulus squre of wave function and the imaginary part of the $d$+$^{208}\mathrm{Pb}$ optical potential. The products for the $l_a=20$, 48 and 55 cases are presented in Figs.~\ref{LiR2} (d), (e) and (f), respectively. The solid lines represent the IAV results, while the dotted lines denote the IAV-cut results with 10 fm cut-offs. It can be observed that for all the three partial waves in Figs.~\ref{LiR2} (d), (e) and (f), value of the product goes to zero rapidly in the region $r>15$ fm, owing to the short-range characteristic of the potential, thus making no contribution to the cross section. As a result, any changes on the wave function outside 15 fm will have no impact on the calculation of NEB cross section. The difference shown in panel (d) is significant; however, the magnitude of the quantity depicted in the figure is small, thus resulting in a relatively negligible NEB cross section. Panels (e) and (f) show no apparent difference between the IAV result and the IAV-cut result for the $l_a=$48 and 55 cases. This observation supports the conclusion that a cut-off on the scattering wave functions $u_{l_a}$ and $u_{l_b}$ does not alter the calculation of NEB cross sections.

\subsection{Application to $(d,pX)$}
On the other hand, we study the deuteron induced inclusive breakup reactions to further investigate the validity of the surface approximation.
First, we carry out the calculation for the $^{208}\mathrm{Pb}(d,p X)$ reaction at $E_{\mathrm{lab}}=70$ MeV.
The proton-target and neutron-target interactions
were adopted from the global parametrization of Koning and Delaroche (KD02)~\cite{koning2003local}. The incoming channel interaction between deuteron and the target is adopted from Ref.~\cite{PhysRevC.74.044615}. For the interaction binding the proton and the neutron in the projectile, we considered the Gaussian form
\begin{equation}
    V(r)=V_0\exp(-r^2/a^2),
\end{equation}
where $a=1.484$ fm, and $V_0$ is fitted to reproduce the experimental binding energy of deuteron.
The model space needed for converged solutions contains partial waves $l\le38$ in the $d$+$^{208}\mathrm{Pb}$ and $p$+$^{209}\mathrm{Pb}^*$ relative
motion, and $l\le18$ in the $n$+$^{208}\mathrm{Pb}$ channel at $E_{\mathrm{lab}}=70$ MeV. 

\begin{figure}[h]
\includegraphics[width=0.48\textwidth]{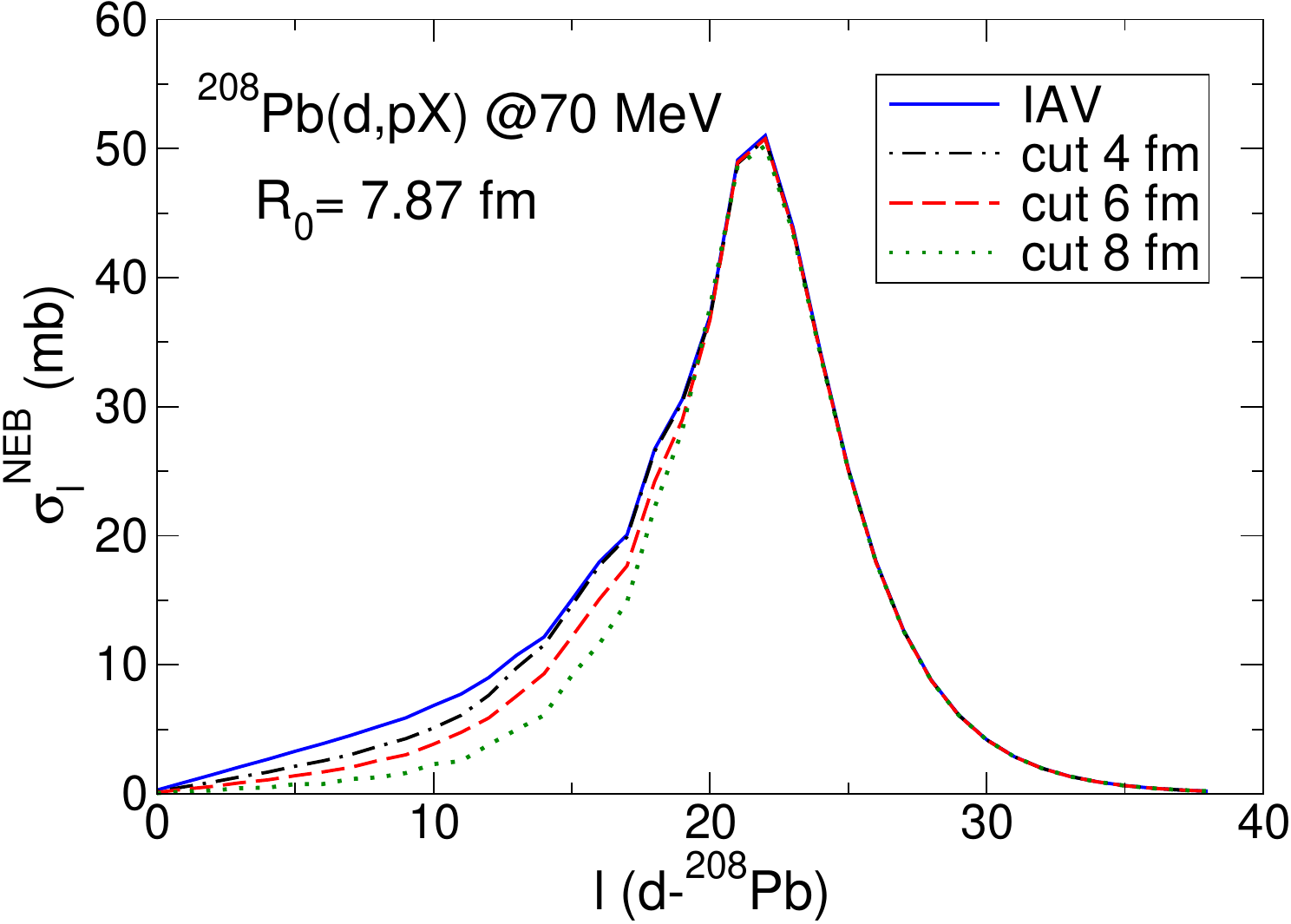}
\caption{Integrated NEB cross section, as a function of the
relative angular momentum between $d$ and $^{208}\mathrm{Pb}$, for the $^{208}\mathrm{Pb}(d,p X)$ reaction at $E_a=70$ MeV. The solid line corresponds to results from the IAV model, while the dot-dashed, dashed, and dotted lines represent the cases for IAV-cut with cut-off radii of 4 fm, 6 fm, and 8 fm, respectively.}
\label{dslx}
\end{figure}

\begin{figure}[h]
\centering
\includegraphics[width=0.48\textwidth]{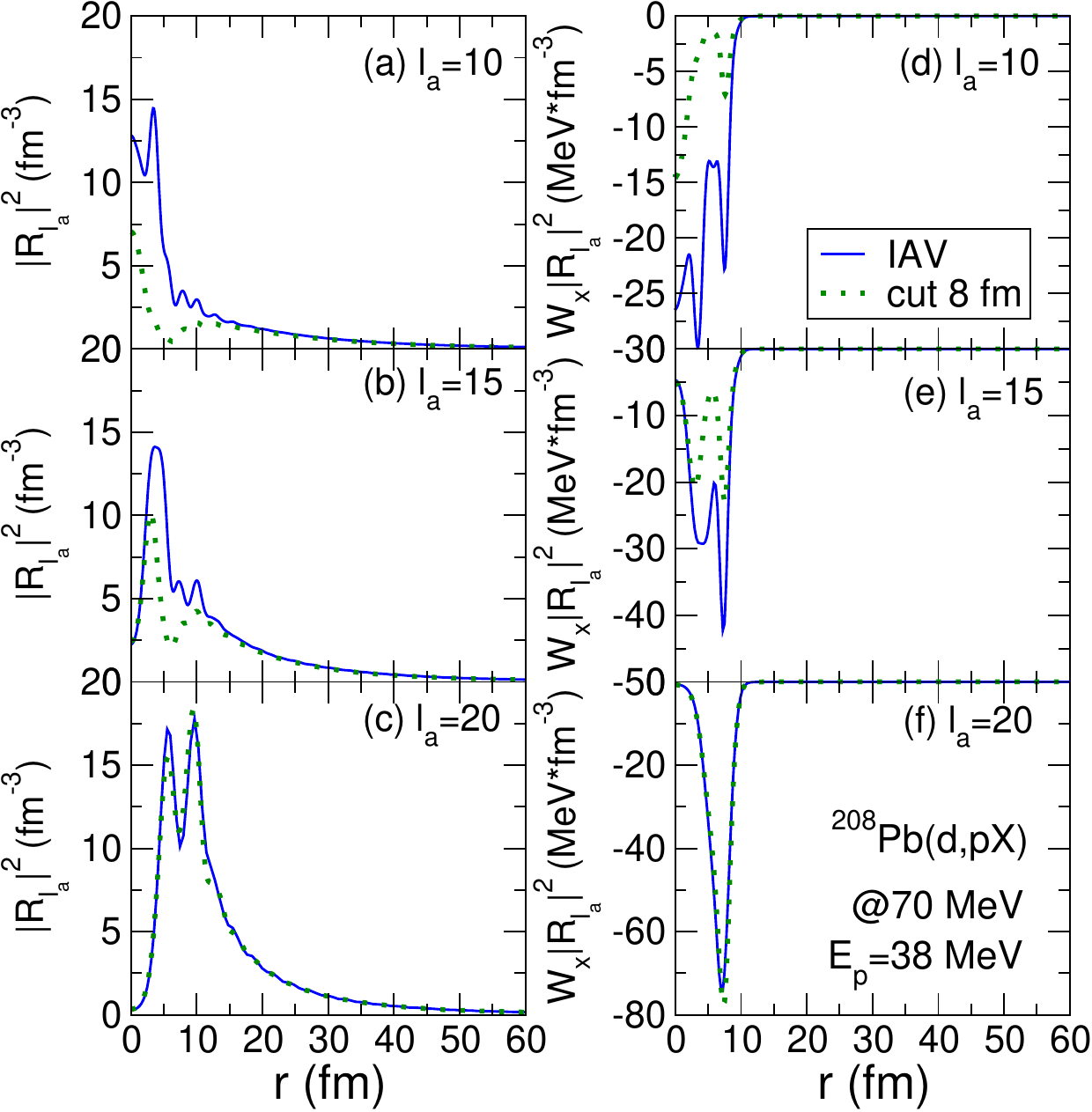}
\caption{The modulus square of  $n$+$^{208}\mathrm{Pb}$ wave function $|R_{l_{a}}|^2$ defined in Eq.~(\ref{eq:rla}) for (a) $l_a=10$, (b) $l_a=15$, and (c) $l_a=20$, for the $^{208}\mathrm{Pb}(d,p X)$ reaction at $E_{\mathrm{lab}}=70$ MeV and relative outgoing kinetic energy $E_{\alpha}=38$ MeV. The solid line represents the IAV result, while the dotted line corresponds to the IAV-cut case with 8 fm cut-off. 
The product of the imaginary part of the $n$+$^{208}\mathrm{Pb}$ potential and the wave function for (d) $l_a=20$, (e) $l_a=48$, and (f) $l_a=55$.}
\label{dR2}
\end{figure}

Similar to the $^6$Li induced cases, here we examine the partial wave dependence of the NEB cross section for this reaction. Fig.~\ref{dslx} presents integrated NEB cross section as a function of $d$+$^{208}\mathrm{Pb}$ relative angular momentum. The solid line corresponds to results from the IAV model, while the dot-dashed, dashed, and dotted lines represent the cases for IAV-cut with cut-off radii of 4 fm, 6 fm, and 8 fm, respectively.
The radius parameter of the imaginary part of the optical potential between $d$ and $^{208}$Pb is $R_0=7.87$ fm~\cite{PhysRevC.74.044615}. 
From this figure, we observed that the difference between lines only occurs in the low partial waves ($l\leq20$). When the cut-off radius increases, the variation in the difference also increases. Compared to the $^6$Li induced reactions, this deuteron induced reaction is more sensitive to inner part of the scattering wave, because the suppression on cross section is enhanced gradually when the cut-off radius increases.
Furthermore, in contrast to previous cases of $^6$Li, no strong absorption effect is observed due to the significant contribution of the low partial wave component to the NEB cross section.
These result illustrates that, without a strong absorption effect, the cut-off of the low partial wave component  will finally manifest in the NEB cross section.

To account for this difference compared to the previous $^6$Li induced cases, we also depicted $|R_{l_a}|^2$ which in this case is the modulus square of the radial part of $n$+$^{208}\mathrm{Pb}$ wave function for $l_a=$ 10, 15, and 20 for this reaction at the outgoing kinetic energy $E_p=$ 38 MeV in CM frame. The results are shown in Fig.~\ref{dR2}.
For the $l_a=10$ and $l_a=15$ cases shown in Figs.~\ref{dR2}(a) and (b), there is a noticeable difference in the wave function obtained from the IAV model and the IAV-cut. Specifically, the wave function obtained from IAV-cut within 10 fm shows a significant reduction. This effect is particularly prominent in the $l_a=10$ case.
This demonstrates that, contrary to the previous situation, the inner part of the $n$+$^{208}\mathrm{Pb}$ wave function with low angular momentum is highly sensitive to the interior part of the scattering functions $u_{l_a}$ and $u_{l_b}$. 
However, in the $l_a=20$ case, there is no clear distinction between the results obtained from the IAV model and the IAV-cut. This is due to the presence of a strong centrifugal barrier, which reduces the scattering wave function inside the barrier to almost zero.

The products of the imaginary part of the $n$+$^{208}\mathrm{Pb}$ potential and modulus square of $n$+$^{208}\mathrm{Pb}$ wave function are also presented in Figs.~\ref{dR2} (d), (e) and (f). Similar to the $^6$Li induced cases, the values of the product go to zero rapidly beyond  the effective range of nuclear force. 
It can be observed from Figs.~\ref{dR2} (d) and (e) that the results obtained from the IAV-cut are significantly suppressed compared to those obtained from the IAV model. This suppression leads to a reduction in the cross section for NEB in the IAV-cut.
Consistent to the results in Fig.~\ref{dslx}, no obvious difference can be seen in Fig.~\ref{dR2} (f) for high partial wave components. These results explain the suppression of cross section in Fig.~\ref{dslx}. Surface approximation for this deuteron induced reaction is not as appropriate as the previous $^6$Li induced cases where the strong absorption effect occurs. Nevertheless, the asymptotic behavior of the wave functions remains unchanged, which indicates that this surface approximation is still valid for EBU calculation.

In the previous $^6$Li induced case, where the IAV-cut and IAV models yield almost identical results, the validity of the surface approximation is supported for both the scattering wave functions $u_{l_a}$ and $u_{l_b}$. However, in the case of the deuteron, it is highly important to investigate whether the cross section suppression observed in the IAV-cut results from the cut-off of both wave functions or only one.
We computed the integrated cross section by exclusively applying a cut-off to either the entrance channel scattering wave function $u_{l_a}$ or the exit channel scattering wave function $u_{l_b}$.
The results are presented in Table~\ref{tab}.
The first column represents the cut-off radius, the second column displays the cross section with a cut-off applied exclusively to the incoming channel wave function $u_{l_a}$, while the third column shows the cross section with a cut-off applied exclusively to the exit channel wave function $u_{l_b}$. The fourth column presents the results obtained by consistently applying cut-offs to both the incoming and exit channel wave functions.

\begin{table}[h]
\begin{tabular}{c|c|c|c}
\hline
\diagbox{cut-off}{$\sigma_\mathrm{NEB}$ (mb)}{method}&  cut $u_{l_a}$  & cut $u_{l_b}$     & cut both   \\
\hline
4~fm &  469   &   470   &  467  \\
6~fm &  449   &   441   &  440  \\
8~fm &  413   &   428   &  410  \\
\hline
\end{tabular}
\caption{Integrated cross section of $^{208}\mathrm{Pb}(d,p X)$ reaction at $E_{\mathrm{lab}}=70$ MeV for different cut-off  radius.} 
\label{tab}
\end{table}
The original IAV result of the integrated cross section is 486 mb.
Table~\ref{tab} shows that the cross sections obtained from cutting $u_{l_a}$, $u_{l_b}$, and cutting both of them are of comparable magnitudes for different cut-off radii. This indicates that the cut-off behaves similarly for both $u_{l_a}$ and $u_{l_b}$, suggesting that no specific cut-off exhibits dominance over another. And applying a cut-off to either $u_{l_a}$ or $u_{l_b}$ will ultimately reduce the cross section when compared to the IAV model.
In another word, both scattering wave functions concurrently contribute to the overall influence on the cross section.
This can be further discussed in a theoretical perspective.
Since the wave functions in the entrance and exit channels are represented in two different sets of Jacobi coordinate, calculating the inhomogeneous 
terms in Eq.(\ref{eq:inhomo}) requires coordinate transformation between these two sets. 
When calculating integrals, the two coordinate variables are not completely orthogonal. 
Therefore, when the product of two scattering wave functions from both the incoming and outgoing channels is integrated, any cut-off applied to one wave function will also exclude the region near the origin of the other wave function. Consequently, the excluded region of the other wave function will not contribute to the final determination of the cross section, regardless of whether this part is cut or not.
In summary, the contribution of the internal parts of the incident and outgoing scattering wave functions cannot be separated when calculating the cross section.

\subsection{\label{dis}Discussion}

Based on the previous calculations and a comparison between the IAV and IAV-cut models, we conclude that the surface approximation is valid for $^6$Li-induced breakup reactions but does not yield satisfactory results for deuteron-induced cases.
In order to further investigate the validity of the surface approximation in the IAV framework, we performed systematic calculations of $^6$Li and deuterons induced inclusive breakup reactions considering different incident energies and target masses. 

We carry out the calculations for 
the $^{28}\mathrm{Si}(d,p X)$ reactions at $E_{\mathrm{lab}}=$2, 6, 10, 20, 30, 60, and 100 MeV,
the $^{208}\mathrm{Pb}(d,p X)$ reactions at $E_{\mathrm{lab}}=$ 20, 30, 50, 70, and 100 MeV,  
the $^{28}\mathrm{Si}(^6\mathrm{Li},\alpha X)$ at $E_{\mathrm{lab}}=$5,  20 ,30, 40, 50, and 100 MeV, 
and $^{208}\mathrm{Pb}(^6\mathrm{Li},\alpha X)$ reactions at $E_{\mathrm{lab}}=$30, 40, 60, 80 and 100 MeV. 
The numerical computation of the NEB cross section using the IAV model are heavy tasks. This is primarily due to the slow convergence of wave functions for many partial waves, the large memory required to store these wave functions, and the need for more grid points to ensure the convergence of the numerical integration. As a consequence, our systematic analysis is computationally intensive and has reached our computing limitations. Thus, we were only able to examine a restricted range of incident energies and target masses, and our conclusions are applicable only under these limited circumstances.

Here we introduce the relative deviation of the integrated NEB cross section to quantify the difference between IAV and IAV-cut:
\begin{equation}
    \delta=\frac{|\sigma_{\mathrm{IAV}}-\sigma_{\mathrm{cut}}|}{\sigma_{\mathrm{IAV}}}\times 100\%,
\end{equation}
where $\sigma_{\mathrm{cut}}$ is the integrated NEB cross section calculated with IAV-cut and $\sigma_{\mathrm{IAV}}$ is result computed directly with the IAV model.

\begin{figure}[h]
\centering
\includegraphics[width=0.48\textwidth]{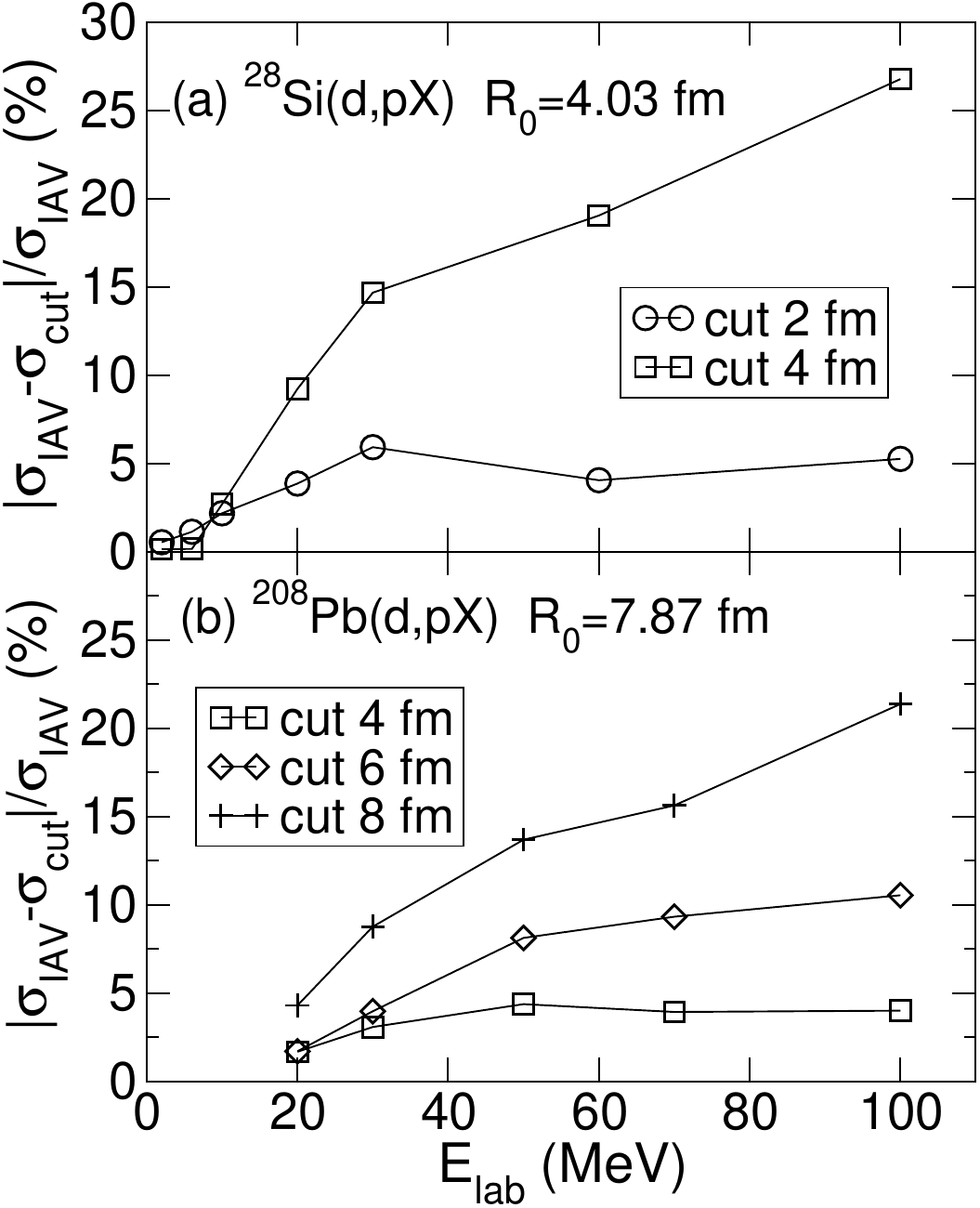} 
 \caption{Relative deviation of integrated NEB cross section for 
 (a) $^{28}\mathrm{Si}(d,p X)$, 
 (b) $^{208}\mathrm{Pb}(d,p X)$.
 Cut-off radii are selected according to the potential parameters, which are included in the figure. 
 Different symbols represent cases with different cut-off radii.  }
 \label{deltad}
 \end{figure}

Figure~\ref{deltad} shows the relative deviation for the deuteron-induced cases discussed above at different incident energies. Panel (a) and (b) correspond to the $^{28}\mathrm{Si}(d,pX)$ and $^{208}\mathrm{Pb}(d,pX)$ cases, respectively. The cut-off radii are selected based on the optical potential parameters and are included in the figures. We use 2 and 4~fm cut-offs for the $^{28}\mathrm{Si}(d,pX)$ reactions and 4, 6, and 8~fm cut-offs for $^{208}\mathrm{Pb}(d,pX)$ reactions. The line with circle, square, diamond, and plus points represents the cases with 2, 4, 6, and 8~fm cut-offs, respectively.
This figure demonstrates an upward trend as the incident kinetic energy increases. With increasing kinetic energy, the relative deviations become more considerable, ranging from a few percent to as much as 25$\%$ in the most extreme scenario. Moreover, the relative deviations still remain moderate for low incident energy near or below the Coulomb barrier, where the strong Coulomb force prevents the wave function from penetrating the interior region. Therefore, setting these wave functions to zero would not affect the calculation of the NEB cross section. Additionally, the figure shows that there is a decrease in the relative deviations with a 2~fm cut-off in panel (a) and a 4~fm cut-off in panel (b). In other words, surface approximation with these cut-offs is still valid in the NEB calculation.

\begin{figure}[h]
\centering
\includegraphics[width=0.48\textwidth]{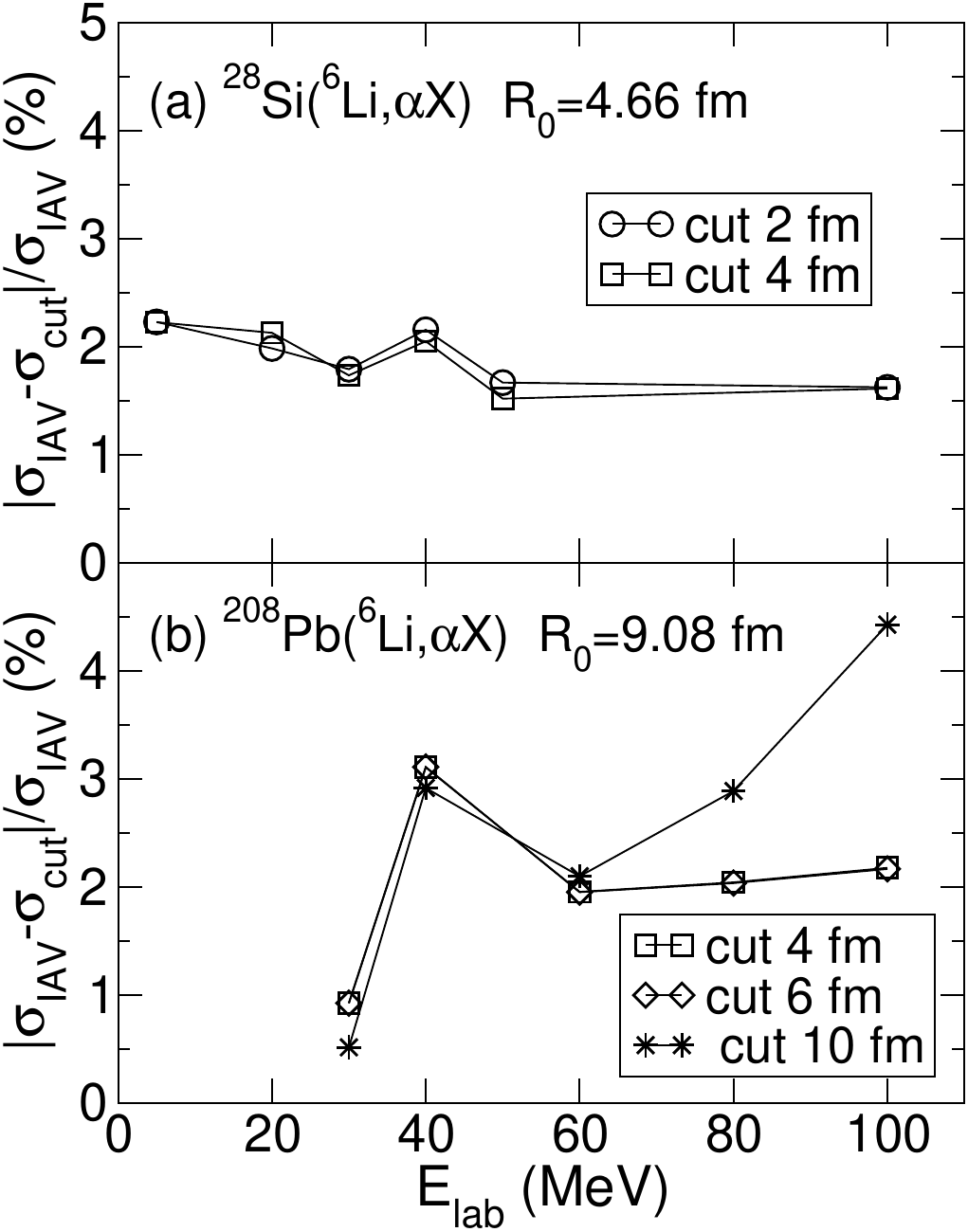} 
 \caption{Relative deviation of integrated NEB cross section for 
 (a) $^{28}\mathrm{Si}(^6\mathrm{Li},\alpha X)$, 
 and (b) $^{208}\mathrm{Pb}(^6\mathrm{Li},\alpha X)$.
 Cut-off radii are selected according to the potential parameters, which are included in the figure. 
 Different symbols represent cases with different cut-off radii.  }
 \label{deltali}
 \end{figure}

Figure~\ref{deltali} illustrates the relative deviation for the $^6$Li induced cases mentioned above at different incident energies.
Panel (a) and (b) present the $^{28}\mathrm{Si}(^6\mathrm{Li},\alpha X)$ and $^{208}\mathrm{Pb}(^6\mathrm{Li},\alpha X)$, respectively. 
We use 2 and 4~fm cut-offs for the $^{28}\mathrm{Si}(^6\mathrm{Li},\alpha X)$ reactions and 4, 6, and 10~fm cut-offs for $^{208}\mathrm{Pb}(^6\mathrm{Li},\alpha X)$ reactions. The lines with circle, square, diamond, and star points represent the cases with 2, 4, 6, and 10~fm cut-offs, respectively.
In Fig.~\ref{deltali} (a), we can observe a decreasing trend in the relative deviation. Specifically, as the kinetic energy increases, the projectile follows a classical trajectory, and a strong absorption effect of the $^6$Li+$^{28}$Si interaction occurs, both of which are key assumptions in semi-classical approaches.
Figures~\ref{deltali}(a) and (b) both demonstrate that the overall relative deviations of the cross section are less than 5$\%$, which is within the typical experimental uncertainty, confirming the validity of surface approximation in these systems. The comparison between Fig.~\ref{deltad} and Fig.~\ref{deltali} reveals that, in the considered circumstances, the surface approximation for $(d,p)$ reactions is not as applicable as it is for $(^6\mathrm{Li},\alpha X)$ reactions. This is evidenced by the deviation values in Fig.~\ref{deltad} being 20\% higher than those in Fig.~\ref{deltali}.

To further investigate the differences caused by $^6$Li and deuteron-induced breakup reactions, we reintroduce the angular dependence of the incoming scattering wave function in order to make a direct comparison with the semi-classical trajectory picture.
The expansion can be written as
\begin{equation}
    \chi_a^{(+)}(\boldsymbol{r})=\sum_l i^l(2l+1)\frac{u_l(r)}{kr}P_l(\cos\theta),
\end{equation}
where the Coulomb phase shift has already been inserted into the radial function $u_l$. 

\begin{figure*}[h]
\centering
\includegraphics[width=1.0\textwidth]{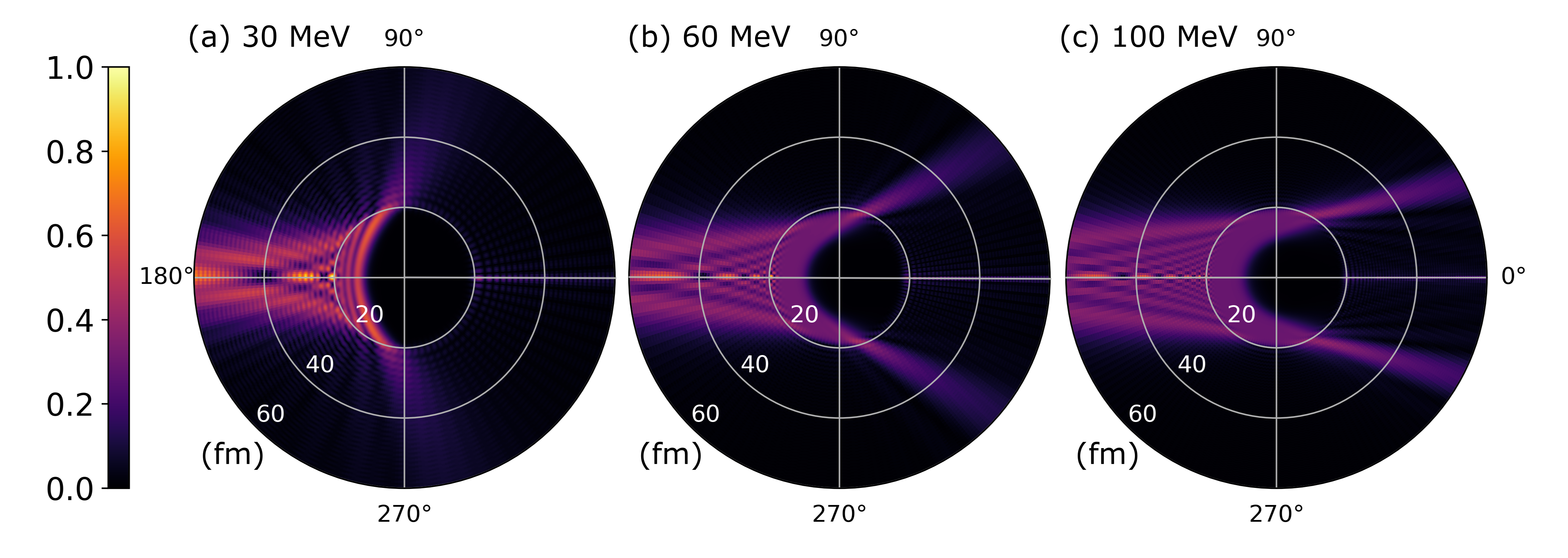} 
\caption{The modulus square of the wave function $|\chi_a^{(+)}|^2$ for $^6\mathrm{Li}$+$^{208}\mathrm{Pb}$ elastic scattering at various bombarding energies. The wave functions are plotted in polar coordinates, with the direction of the incoming projectile being $0^{\circ}$. The three panels correspond to (a) $E_{a}=30$ MeV, (b) $E_{a}=60$ MeV, and (c) $E_{a}=100$ MeV.}
\label{fig:lipolar}
\end{figure*}

Figure~\ref{fig:lipolar} depicts the modulus square of the wave function $|\chi_a^{(+)}|^2$ in the x-z plane for the elastic scattering of $^6\mathrm{Li}$+$^{208}\mathrm{Pb}$ reaction at three different bombarding energies. This figure is arranged as a heatmap in polar coordinate, where lighter colors represent larger probability of finding a particle according to the probability interpretation of wave function. At 30 MeV, $^6$Li is incapable of penetrating $^{208}$Pb and instead is diffracted before reaching the target. A clearly defined classical trajectory becomes apparent in Figs.~\ref{fig:lipolar}(b) and (c), because
as the scattering angle decreases, the peak of probability density forms a straight line.
Additionally, Figs.~\ref{fig:lipolar}(b) and (c) demonstrate that the probability of detecting a particle at forward region is exceptionally low because there are no bright points in the forward region. This lack of probability in the forward region demonstrates the strong absorption effect of the $^6$Li+$^{208}$Pb interaction, in which low angular momentum components are fully fused into the target and do not contribute to the NEB cross section.  These results affirms the validity of introducing radial cut-offs at high incident energies in IAV framework.

\begin{figure*}[h]
\centering
\includegraphics[width=1.0\textwidth]{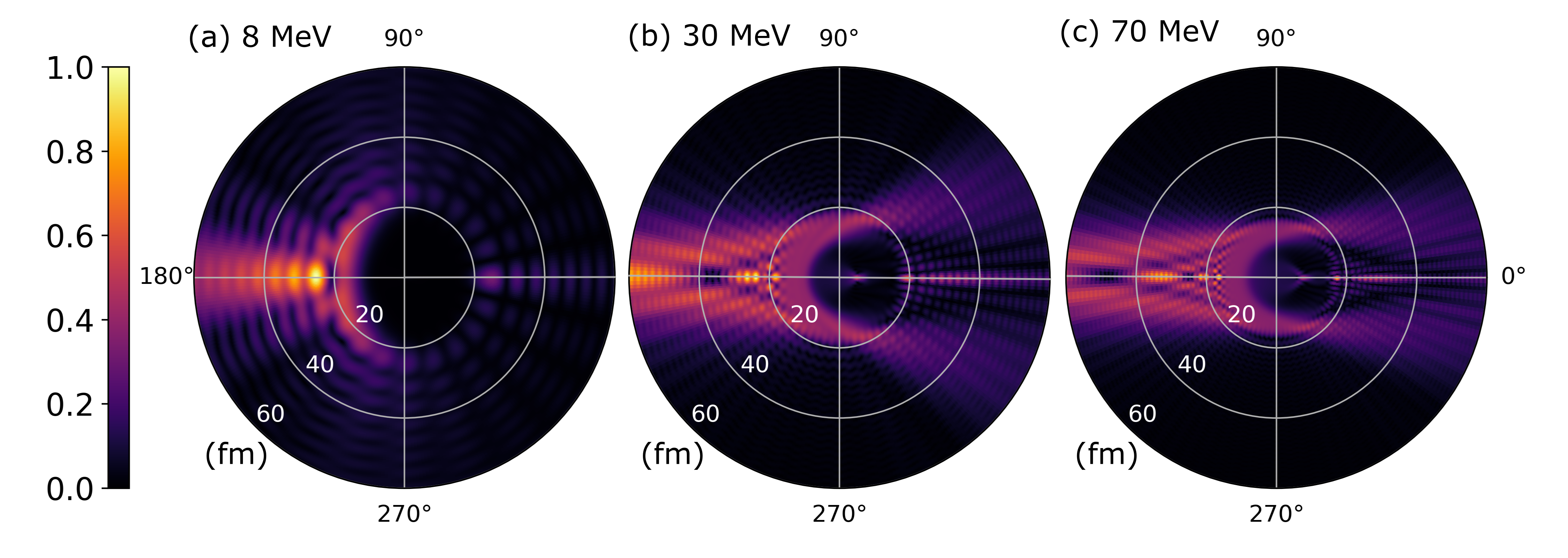} 
\caption{The modulus square of the wave function $|\chi_a^{(+)}|^2$ for 
$d$+$^{208}\mathrm{Pb}$ elastic scattering at various bombarding energies. The wave functions are plotted in polar coordinates, with the direction of the incoming projectile being $0^{\circ}$. The three panels correspond to (a) $E_{a}=8$ MeV, (b) $E_{a}=30$ MeV, and (c) $E_{a}=70$ MeV.}
\label{fig:dpolar}
\end{figure*}

As a comparison, the wave function for the elastic scattering of $d$+$^{208}$Pb is depicted in Fig.~\ref{fig:dpolar}. Unlike the previous case, there is a strong interference in the forward angle in all the panels of Fig.~\ref{fig:dpolar}, showing strong wave-like characteristics. This strong distortion of the scattering wave functions lack a correspondence to the classical trajectory picture, and thus the surface approximation fails. Besides, there is no evidence of a strong absorption effect of the $d$+$^{208}$Pb interaction, because there are many light points in the forward angle. 

Comparing wave functions of these reactions makes it straightforward to establish the validity of surface approximation based on whether the incoming channel elastic scattering process has a clear correspondence to a classical trajectory.

\section{\label{con}Conclusion}
We present a study on the nonelastic breakup reactions induced by weakly bound nuclei with a fully quantum-mechanical model from Ichimura, Austern, and Vincent. Corresponding to the classical picture of trajectory in reaction processes, we introduce a radial cut-off to investigate the validity of the surface approximation on a fully quantum-mechanical basis. With a proper selection of the cut-off radius, we apply this surface approximation to 
the $(^6\mathrm{Li},\alpha X)$ and $(d,p X)$ reactions. 

We observed that the approximated cross sections computed with cut-offs for the $(^{6}\mathrm{Li},\alpha X)$ reactions exhibit good overall agreement with the accurate calculations. These results indicate that NEB cross section is insensitive to the inner wave function, and the semi-classical picture is valid. For $(d,pX)$ reactions, a non-negligible loss of cross-sections was observed after the cut-off, suggesting a strong dependence on the inner wave functions at low energies in the IAV framework. Setting the inner wave function to zero in $^{6}\mathrm{Li}$ induced reactions has little effect on the cross-section due to a strong absorption of small angular momentum components in the entrance channel. However, in the case of deuteron induced reactions, the distortion caused by the nuclear and the Coulomb forces does not correspond to a semi-classical trajectory picture. Consequently, there is a relatively stronger dependence on the inner scattering wave function in the IAV framework.

However, these conclusions are based on a very limited set of systems. Further studies involving higher incident energies and more targets are called for. 
We plan to optimize our computer code so that we can conduct calculations for reactions involving heavier targets and higher energy, which are currently beyond our computing capability.

\begin{acknowledgments}
This work has been supported by National Natural Science Foundation of China (Grants No.12105204, No.12035011, and No.11975167), by the Fundamental Research Funds for the Central Universities.
\end{acknowledgments}

\bibliography{SA.bib}
\end{document}